
%
%
\def\npix{{N_{pix}}}
\def\section#1 {\bigskip\centerline{\bf #1.}\medskip}
\def\etal{{\it et al.\/}}
\def\muk{{\mu {\rm K}}}
\def\half{\hbox{$1\over 2$}}
\magnification=\magstep1
\baselineskip 16pt
\line{}
\centerline{\bf Wiener Filtering of the {\sl COBE} DMR Data.}
\vskip 0.6 truein
\centerline{Emory F. Bunn$^{1,2}$, Karl B. Fisher$^3$, Yehuda Hoffman$^4$,}
\centerline{Ofer Lahav$^3$, Joseph Silk$^1$, and Saleem Zaroubi$^4$}
\vskip 0.5 truein
{\sl
\parindent=0pt
\hangindent=12pt
1. Departments of Physics and Astronomy and Center for
Particle Astrophysics,
University of California, Berkeley CA  94720

2. E-mail:  {\rm bunn@pac2.berkeley.edu}

3. Institute of Astronomy, Madingley Rd., Cambridge CB3 OHA,
UK

4. Racah Institute of Physics, The Hebrew University,
Jerusalem, Israel
}
\vskip 0.5 truein
\centerline{\bf Abstract.}
\vskip 0.3 truein
We derive an optimal linear filter to suppress the noise from the {\sl
COBE} DMR sky maps for a given power spectrum.  We then apply the
filter to the first-year DMR data, after removing pixels within
$20^\circ$ of the Galactic plane from the data.  The filtered data have
uncertainties 12 times smaller than the noise level of the raw data.
We use the formalism
of constrained realizations of Gaussian random fields to assess the
uncertainty in the filtered sky maps.  In addition to improving the
signal-to-noise ratio of the map as a whole, these techniques allow
us to recover some information about the CMB anisotropy in the missing
Galactic plane region.  From these
maps we are able to determine which hot and cold spots in the data are
statistically significant, and which may have been produced by noise.  In
addition, the filtered maps can be used for comparison with other
experiments on similar angular scales.
\vfill
{\noindent Preprint Number CfPA--TH--94--20.  Submitted to
{\sl Astrophysical Journal Letters.}}
\eject

\section{Introduction}
The discovery by the {\sl COBE} DMR of large-scale anisotropy in the cosmic
microwave background (CMB) has ushered in a new, more quantitative era
in the study of microwave background anisotropy (Smoot \etal~1992).
The statistical
properties of the fluctuations in the DMR sky maps provide an
important estimate of the amplitude of cosmological fluctuations on
very large scales, which are impossible to observe in any other
way.

Most analyses of the DMR sky maps have focused on the problem of
estimating statistical properties of the anisotropy such as the power
spectrum ({\it e.g.\/} Wright \etal~1992, Seljak \& Bertschinger 1993, Wright
\etal~1994), or on testing the hypothesis that the anisotropy
obeys Gaussian statistics ({\it e.g.\/}
Hinshaw \etal~1994, Smoot \etal~1994).  In addition, statistical comparisons
have been made between the DMR sky maps and data from other experiments
(Watson \etal~1993, Ganga \etal~1993).  Less attention has been paid
to studying
the spatial properties of the sky maps themselves to determine, for
example, which hot and cold spots on the maps are likely to be real,
and which are dominated by noise.  Such an analysis has the
potential to be quite useful in comparing the DMR results to other
experiments that probe comparable angular scales, such as the FIRS and
Tenerife experiments (Meyer \etal~1991, Hancock \etal~1994).

There are two major difficulties in studying the DMR maps.  The
signal-to-noise ratio in each pixel is quite low, and pixels near the
Galactic plane are contaminated by Galactic emission.  These problems
place strict limits on the accuracy with which the true structure can be
recovered (Bunn, Hoffman, \& Silk 1993)
In this {\sl
Letter} we apply the techniques of Wiener filtering and constrained
realizations in an attempt to mitigate these problems.  We assume the
correctness of the canonical theory of large-scale CMB anisotropy:
that the anisotropy forms a Gaussian random field with a power
spectrum that is a power law in spatial wavenumber $k$.  We then remove
the pixels that are presumed to be contaminated, and apply
an optimal linear filter to the data in an attempt to clean up the
noise and see the underlying structure.  The technique of constrained
realizations helps quantify the uncertainties associated with this method.

\section{The Wiener Filter and Constrained Realizations}
The Wiener filtering described in this work is similar to a technique applied
recently to galaxy catalogues
(Lahav \etal~1994).
The formalism for making constrained realizations of Gaussian random
fields has also been discussed elsewhere (Hoffman \& Ribak 1991, Gannon \&
Hoffman 1993),
as has the connection between
this formalism and the Wiener filter (Zaroubi \etal~1994).
In this section we simply present
these techniques in a language appropriate for analyzing the DMR data.

Let us begin by writing the true anisotropy
$\Delta T$ as an expansion in spherical harmonics\footnote{*}
{Throughout this {\sl Letter}, the symbol $Y_{lm}$ will denote a real-valued
spherical harmonic.
The real-valued spherical harmonics are
simply the conventional spherical harmonics with complex exponentials
replaced by ordinary trigonometric functions:
$$
Y_{lm}(\theta,\phi)=\cases{Y_{lm}^{(conv)}(\theta,\phi)& for $m=0$\cr
Y_{lm}^{(conv)}(\theta,0)\sqrt{2}\cos m\phi& for $1\le m\le l$\cr
Y_{lm}^{(conv)}(\theta,0)\sqrt{2}\sin m\phi& for $-l\le m\le -1$}.
$$
These functions satisfy the usual orthonormality condition $\int Y_{lm}(\theta,
\phi)Y_{l'm'}(\theta,\phi)\,d\Omega=\delta_{ll'}\delta_{mm'}$.
}:
$$
\Delta T({\bf\hat r})=\sum_{l=2}^\infty\sum_{m=-l}^l
a_{lm}Y_{lm}({\bf\hat r}).\eqno{(1)}
$$
We assume that the anisotropy forms a Gaussian random field: each
coefficient $a_{lm}$ is an independent Gaussian random variable of
zero mean ({\it e.g.\/} Bond \& Efstathiou 1987).
The angular power spectrum of this Gaussian random field
is given by
$$
C_l=\langle a_{lm}^2\rangle,\eqno{(2)}
$$
where the angle brackets denote an ensemble average.

Each DMR pixel contains a measurement of the beam-smoothed temperature
at a particular point:  if we denote the $i$th data point by $D_i$, then
$$
D_i=\left(\Delta T\star B\right)({\bf\hat r}_i)+n_i.\eqno{(3)}
$$
Here $B$ denotes the DMR beam pattern, the star represents a convolution,
${\bf\hat r}_i$ is a unit vector in the direction of the $i$th pixel, and
$n_i$ is the noise in the pixel.  The convolution is performed quite simply
in spherical harmonic space:  the coefficients  of the spherical
harmonic expansion of $(\Delta T\star B)$ are given by $a'_{lm}=B_l a_{lm}$,
where
$$
B_l=2\pi\int_0^\pi B(\theta)P_l(\cos\theta)\sin\theta\,d\theta,\eqno{(4)}
$$
and $P_l$ is a Legendre polynomial.
The beam pattern is often approximated by
a Gaussian, but we use the actual DMR beam pattern given in Wright
\etal~(1994).

The noise is presumed to consist of
independent Gaussian random numbers with known variances $\sigma_i^2$.
The assumption that the noise in different pixels is uncorrelated has
been recently confirmed by Lineweaver \etal~(1994).  This represents
one significant difference between the present work and the analysis
of galaxy catalogues in Lahav \etal~(1994).  The shot noise in the galaxy
distribution is not independent from pixel to pixel, because it has been
smoothed along with the signal.  The DMR instrumental noise enters the data
after the smoothing by the beam has already taken place.

Pixels near the Galactic plane are presumed to be contaminated.  Even
though we work with the ``Reduced Galaxy'' linear combination of
sky maps (Smoot \etal~1992),
 the pixels within $20^\circ$ of the Galactic plane are
excised from the data set.  The number of pixels left after this
operation is $\npix=4038$.

If we assume that we know the power spectrum $C_l$, we can apply a Wiener
filter to reconstruct $\Delta T$ from the data.  We begin by generating
initial estimates $b_{lm}$ of the spherical harmonic coefficients:
$$
b_\mu=\sum_{i=1}^\npix w_i Y_\mu({\bf\hat r}_i) D_i,\eqno{(5)}
$$
where the weights are given by $w_i=1/\sigma_i^2$.
Here and hereafter, Greek indices denote pairs $(l,m)$.  (The estimated
coefficients $ b_\mu$ are not normalized properly, but the overall
normalization is irrelevant for our purposes.)  The monopole and dipole
terms are removed from the DMR sky maps before these recovered coefficients
are computed.

The estimated spherical harmonic coefficients
$b_\mu$ are related to the true coefficients $a_\mu$ as follows:
$$
b_\mu=\sum_{\nu}W_{\mu\nu}B_{\nu}a_{\nu}+\sum_{i=1}^\npix
 w_i Y_\mu({\bf\hat r}_i)n_i,\eqno{(6)}
$$
where
$$
W_{\mu\nu}=\sum_{i=1}^\npix w_i Y_\mu({\bf\hat r}_i)Y_\nu({\bf\hat r}_i).
\eqno{(7)}
$$
The matrix $W$ describes the coupling that is introduced
by incomplete sky coverage
between the different spherical harmonics (Peebles 1980, Scharf \etal~1992,
Lahav~\etal 1994).

We want to apply a linear filter to these coefficients to get a vector of
new coefficients
$\vec c$ that is as close as possible to the original coefficients $\vec a$,
in the sense of least squares.  That is to say, we want to define a set
of coefficients
$$
\vec c=F\vec b,\eqno{(8)}
$$
where the matrix $F$ has been chosen to minimize
$$
\Delta=\left<|\vec c-\vec a|^2\right>=\left<\sum_\mu(c_\mu-a_\mu)^2
\right>\eqno{(9)}
$$
By substituting equations (5), (6) and (8) into this expression, and setting
$\partial\Delta/\partial F_{\mu\nu}=0$, we find the appropriate filter $F$:
$$
F= CB(WBCB+1)^{-1}.\eqno{(10)}
$$
In the above equation $W$ is the matrix with elements $W_{\mu\nu}$,
and $B$ and $C$ are diagonal matrices representing the power spectrum
and the beam pattern: $C_{\mu\nu}=C_l\delta_{\mu\nu}$ and $B_{\mu\nu}=B_l
\delta_{\mu\nu}$, where $l$ is
the multipole number corresponding to $\mu$.

Once the Wiener-filtered coefficients $\vec c$ have been determined, the
reconstructed temperature anisotropy at any point on the sky is simply
$$
\Delta T_W({\bf\hat r})=\sum_\mu c_\mu Y_\mu({\bf\hat r}).\eqno{(11)}
$$

Of course, we do not expect the temperature at any particular point on the
sky to match the Wiener reconstructed value exactly.  We can assess the
expected fluctuations about the reconstructed map with the formalism
of constrained realizations.
In deriving the Wiener filter, we assumed that we knew only the power spectrum
$C_l$ ({\it i.e.,} the variances of the $a_{lm}$'s).  We have not yet used
our additional assumption that the
probability distribution function for the $a_{lm}$'s is Gaussian.
With this assumption,
we can use the formalism of conditional probability to
compute the probability of any particular
realization $\Delta T({\bf\hat r})$ being the true anisotropy:  if the
coefficients
of the spherical harmonic expansion of $\Delta T$ are $a_{lm}$, then
the probability of $\Delta T$ being the true anisotropy, given the data $D$,
is
$$\eqalignno{
p(\Delta T|D)&\propto\, p(\Delta T)\,p(D|\Delta T)\cr
&\propto\,
\exp(-\half\vec a^TC^{-1}\vec a)\,\exp\left(-{1\over2}\sum_{i=1}^\npix {(D_i-
(\Delta T\star B)
({\bf\hat r}_i))^2\over\sigma_i^2}\right).&(12)
}$$
The first term is simply the probability of getting the realization
$\Delta T$ given the power spectrum, and the second term is the
probability of getting the observed data given the underlying field
$\Delta T$.  If we expand $\Delta T$ in terms of the $a_{lm}$'s, and
complete the square in the exponential, this expression becomes
$$
p(\Delta T|D)\propto\,\exp\left(-\half(\vec a-\vec c)^T
F^{-1}(\vec a-\vec c)\right).
\eqno{(13)}
$$
This is a Gaussian probability distribution with mean
$\vec c$ and covariance matrix $F$ (Rybicki \& Press 1992).  This probability
distribution takes its maximum value when $\Delta T=\Delta T_W$.

{}From this information we can easily compute the variance of
$\Delta T({\bf\hat r})$
about the mean value $\Delta T_W({\bf\hat r})$, which we can interpret as the
uncertainty in the Wiener reconstructed anisotropy at the point ${\bf\hat r}$.
In addition, we can make constrained realizations of the anisotropy $\Delta T$
with the probability distribution (13).

\section{Results}
We applied the Wiener filter to the DMR first-year ``Reduced Galaxy''
sky map, shown in Figure 1.
We made three different choices for the power spectrum,
corresponding to power-law indices $n=0.5,1,1.5$ for the primordial
matter power spectrum.  The normalization in each case was given by
the correlation function analysis of Seljak \& Bertschinger (1993):
The quadrupole
amplitude $Q_{rms-PS}=15.7\,
\exp(0.46(1-n))\,\muk$.  The second-year DMR data are consistent
with this amplitude, although the data prefer a slightly lower normalization
(Bennett \etal~1994)

Given $n$ and $Q_{rms-PS}$,
the power spectrum $C_l$ is
$$
C_l=\left(4\pi Q_{rms-PS}^2\over 5\right){\Gamma\left(2l+n-1\over 2\right)
\Gamma\left(9-n\over 2\right)\over\Gamma\left(2l+5-n\over 2\right)
\Gamma\left(3+n\over 2\right)}.\eqno{(14)}
$$
(Bond \& Efstathiou~1987).

The filtered sky maps for these three power spectra are shown in Figure 2.
Multipoles with $l \le 30$ were included in the filtering, and the
final results were quite insensitive to variations in this cutoff.
The insensitivity to the cutoff results from the high noise level in
the data:  modes with large $l$ are dominated by noise, and so are
highly suppressed by the Wiener filter.  This is in contrast with the
results of Lahav \etal~(1994) for the galaxy distribution,
which have a higher signal-to-noise
ratio, and are therefore more sensitive to the cutoff.

The three maps show similar qualitative features, but the maps with
larger values of the spectral index $n$ have more small-scale structure.
This is not surprising:  it simply reflects the fact that the assumed
power spectrum $C_l$ has more small-scale power for large $n$.

We know that the difference between the actual anisotropy and the
Wiener-filtered map is a Gaussian random field with covariance matrix
$F$.  The uncertainty $U({\bf\hat r})$ in the reconstructed map at any
particular point
$\bf\hat r$ on the sky is simply the square root of the variance of
this field at $\bf\hat r$.  The r.m.s value of $U({\bf\hat r})$ over
the region outside of the galactic cut is $20\,\muk$, which is
lower than the $240\,\muk$ r.m.s. pixel noise in the raw data.
(This is what we mean when we say that the Wiener filter ``suppresses
noise.'')  However, the filtering process also suppresses some of the
signal.  In order to compare the filtered maps to the raw data, we define
a signal-to-noise ratio for the filtered maps as $\Delta T_W(
{\bf\hat r})/U({\bf\hat r})$.

Figure 3 shows contour plots of this
signal-to-noise ratio for the three filtered maps.  For the Harrison-Zel'dovich
map, 308
of the 4038 pixels that lie outside of the $20^\circ$ Galactic
cut have signal-to-noise ratios greater than 2.  This is a
considerable improvement over the 211 pixels in the raw data that
have signal-to-noise ratios greater than 2.  Since no data were used
from within the Galactic cut region, the area near the Galactic
equator is reconstructed with somewhat lower significance.

The signal-to-noise map in Figure 3 contains no information about
correlations between the fluctuations at different points in the sky:
it was made using only the diagonal part of the covariance matrix.  We
can get a more complete picture of the expected fluctuations about the
Wiener-filtered map by making constrained realizations.  Figure 4
shows three constrained realizations of the CMB anisotropy.  These
maps were made with the assumption of a Harrison-Zel'dovich power
spectrum, and have been smoothed with the {\sl COBE} beam pattern.  The
constrained realizations show a number of features that persist from
map to map.  In particular, the hot spots near the Galactic
coordinates $(l,b)=(275^\circ,-36^\circ)$ and $(55^\circ,65^\circ)$
appear highly
significant, as does the large cold region around $(260^\circ,50^\circ)$.

\section{Conclusions}
Wiener filtering is a promising tool for the analysis of
CMB sky maps.  Filtering provides a significant improvement in
the signal-to-noise ratio in the regions covered by the raw data, and
allows some information to be reconstructed about the anisotropy within
the Galactic cut region.
It is possible to identify several hot and cold spots in the map which
carry high statistical significance.

This reduction in noise is not without a price.  In order to apply the
Wiener filter, one needs to assume a power spectrum.  The cleaned data
therefore depend on more assumptions than do the raw data.  However,
the maps do not undergo great qualitative changes as one varies the
slope of the power spectrum over a wide range of reasonable values.

The expected fluctuations about the Wiener filtered maps are assessed
with constrained realizations.  This technique can be used to perform
Monte-Carlo simulations of the CMB for comparison with experiments on
similar angular scales such as the Tenerife (Hancock \etal~1994)
and FIRS (Ganga \etal~1993) experiments.  In particular, Hancock
\etal~(1994) have reported a significant fluctuation at a particular
location on the sky.  Due to the differencing technique of the Tenerife
experiment and the difference in angular scales, it is fruitless to try
to identify this feature by eye on the filtered DMR maps; however,
a quantitative comparison of the Tenerife and FIRS data sets with
the filtered DMR maps may prove revealing.  This approach makes use
of both amplitude and phase information in the two data sets, and is
therefore potentially more powerful than power-spectrum estimation techniques,
which in general throw away the phase information.

\section{Acknowledgements}
The {\sl COBE} data sets were developed by the NASA
Goddard Space Flight Center under the guidance
of the {\sl COBE} Science Working Group and were
provided by the NSSDC.
This research has been supported in part by a grant from NASA.
Y.H. acknowledges the hospitality of the Center for Particle Astrophysics
and the Department of Astronomy at U.C. Berkeley.

\vfill\eject
\parindent=0pt
\parskip=12pt
\centerline{\bf Figure Captions}

Figure 1.  The DMR Reduced Galaxy sky map is shown.  The upper panel
shows the raw data.  The lower panel is the data after smoothing with
a $7^\circ$ FWHM Gaussian.  Both maps are Aitoff projection, with the
North Galactic pole $b=90^\circ$ at the top and the Galactic center
$(l,b)=(0^\circ,0^\circ)$ at the center of the plot.  Longitude
increases from right to left.  The range of temperatures covered in the
map is $(-1000\,\muk,1100\,\muk)$.

Figure 2.  The Wiener-filtered DMR Reduced Galaxy maps are shown for
three different choices of initial power spectrum.  The spectral index
$n$ and quadrupole normalization $Q_{rms-PS}$ take values
$(n,Q_{rms-PS})=(1.5,12.5\,\muk), (1.,15.7\,\muk),(0.5,19.8\,\muk)$ in
the three panels.  The range of temperatures covered in the map is
$(-60\,\muk,65\,\muk)$.  The maps are in Galactic coordinates as in
Figure 1.

Figure 3.  Contour plots are shown of the signal-to-noise ratio in the
Wiener-filtered maps.  The contours show regions in which the
reconstructed temperature $\Delta T_W$ is 1, 2, and 3 times the noise
level.  Dotted contours denote regions where $\Delta T_W<0$.  The maps
are in Galactic coordinates as in Figure 1.

Figure 4.  Three constrained realizations of the DMR sky map are
shown.  The maps were made with the assumption of a
Harrison-Zel'dovich ($n=1$) power spectrum, including multipoles up to
$l_{max}=30$.  The maps shown were smoothed with the DMR beam pattern.
The range of temperatures covered in the maps is
$(-122\,\muk,106\,\muk)$.  The maps are in Galactic coordinates as in
Figure 1.

\vfill\eject
\parindent 0pt
\parskip 10pt
\baselineskip=\normalbaselineskip
\centerline{\bf References.}
\bigskip\bigskip\bigskip

Bennett, C.L. \etal~1994, Ap. J., submitted.

Bond, J.R. \& Efstathiou, G. 1987, M.N.R.A.S., 226, 655.

Bunn, E.F., Hoffman, Y. \& Silk, J. 1994, Ap. J., in press.

Ganga, K., Cheng, E., Mayer, S., \& Page, L. 1993, Ap. J. Lett., 410, L57.

Ganon, G. \& Hoffman, Y. 1993, Ap. J. Lett., 415, L5.

Hancock, S. \etal~1994, Nature, 367, 333.

Hinshaw, G., Kogut, A., Gorski, K.M,, Banday, A.J., Bennett, C.L.,
Lineweaver, C., Lubin, P., Smoot, G.F., \& Wright, E.L.  1994, Ap.J.,
submitted ({\sl COBE} Preprint 93-12).

Hoffman, Y. \& Ribak, E. 1991, Ap. J. Lett. 380, L5.

Lahav, O., Fisher, K.B., Hoffman, Y., Scharf, C.A., \& Zaroubi, S. 1994, Ap.
J. Lett., 423, L93.

Lineweaver, C.H., Smoot, G.F., Bennett, C.L., Wright, E.L., Tenorio, L.,
Kogut, A., Keegstra, P.B., Hinshaw, G., \& Banday, A.J. 1994, Ap. J.,
submitted.

Meyer, S.S., Cheng, E.S., \& Page, L. 1991, Ap. J. Lett., 371, L7.

Rybicki, G.B. \& Press, W.H.  1992, Ap. J. 393, 169.

Scharf, C.A., Hoffman, Y., Lahav, O., \& Lynden-Bell, D.~1992, M.N.R.A.S.,
254,389.

Seljak, U. \& Bertschinger, E. 1993, Ap. J. Lett., 417, L9.

Smoot, G.F. \etal~1992, Ap. J. Lett., 396, L1.

Smoot, G.F., Tenorio, L., Banday, A.J., Kogut, A., Wright, E.L., Hinshaw, G.,
\& Bennett, C.L. 1994, Ap.J., submitted.

Watson, R.A. \& Gutierrez de la Cruz, C.M. 1993, Ap. J. Lett. 419, L5.

Wright, E.L. \etal 1992, Ap. J. Lett., 396, L13.

Wright, E.L. \etal 1994, preprint.

Zaroubi, S., Hoffman, Y., Fisher, K.B., Lahav, O., \& Lynden-Bell, D. 1994,
in preparation.

\bye